\documentclass[sigconf]{acmart}
\usepackage{graphicx}
\usepackage{subcaption}

\AtBeginDocument{%
  \providecommand\BibTeX{{%
    \normalfont B\kern-0.5em{\scshape i\kern-0.25em b}\kern-0.8em\TeX}}}

\setcopyright{acmcopyright}
\copyrightyear{2023}
\acmYear{2023}

\acmConference[Fast Machine Learning for Science  ICCAD, 2023]{}{November 2, 2023}{San Francisco, CA}
\begin{document}

\title{FPGA Deployment of LFADS for Real-time Neuroscience Experiments }

\author{Xiaohan Liu}
\email{xliu1626@uw.edu}
\affiliation{%
  \institution{University of Washington}
  \city{Seattle}
  \postcode{WA 98195}
  \country{USA}
}

\author{ChiJui Chen}
\email{silencekugel.ee05@nycu.edu.tw}
\affiliation{%
  \institution{National Yang Ming Chiao Tung University}
   \city{Hsinchu}
   \country{Taiwan}
}

\author{YanLun Huang}
\email{yanlun172@gmail.com}
\affiliation{%
  \institution{National Yang Ming Chiao Tung University}
   \city{Hsinchu}
   \country{Taiwan}
}

\author{LingChi Yang}
\email{hisky1256.ee11@nycu.edu.tw}
\affiliation{%
  \institution{National Yang Ming Chiao Tung University}
   \city{Hsinchu}
   \country{Taiwan}
}

\author{Elham E Khoda}
\email{ekhoda@uw.edu}
\affiliation{%
  \institution{University of Washington}
  \city{Seattle}
  \postcode{WA 98195}
  \country{USA}
}

\author{Yihui Chen}
\email{yihuic@uw.edu}
\affiliation{%
  \institution{University of Washington}
   \city{Seattle}
   \country{USA}
}

\author{Scott Hauck}
\email{hauck@uw.edu}
\affiliation{%
  \institution{University of Washington}
  \city{Seattle}
  \postcode{WA 98195}
  \country{USA}
}

\author{Shih-Chieh Hsu}
\email{schsu@uw.edu}
\affiliation{%
  \institution{University of Washington}
  \city{Seattle}
  \postcode{WA 98195}
  \country{USA}
}

\author{Bo-Cheng Lai}
\email{bclai@nycu.edu.tw}
\affiliation{%
  \institution{National Yang Ming Chiao Tung University}
   \city{Hsinchu}
   \country{Taiwan}
}

\renewcommand{\shortauthors}{Liu, et al.}


\begin{abstract}
Large-scale recordings of neural activity are providing new opportunities to study neural population dynamics.
A powerful method for analyzing such high-dimensional measurements is to deploy an algorithm to learn the low-dimensional latent dynamics. 
LFADS (Latent Factor Analysis via Dynamical Systems) is a deep learning method for inferring latent dynamics from high-dimensional neural spiking data recorded simultaneously in single trials.
This method has shown a remarkable performance in modeling complex brain signals with an average inference latency in milliseconds. 
As our capacity of simultaneously recording many neurons is increasing exponentially, it is becoming crucial to build capacity for deploying low-latency inference of the computing algorithms.  
To improve the real-time processing ability of LFADS, we introduce an efficient implementation of the LFADS models onto Field Programmable Gate Arrays (FPGA).
Our implementation shows an inference latency of 41.97 $\mu$s for processing the data in a single trial on a Xilinx U55C.

\end{abstract}

\keywords{neuroscience, latent variable model, recurrent neural network, behavioural neuroscience, \texttt{hls4ml}, FPGA}

\maketitle

\section{Introduction}

\begin{figure*}[!ht]
     \centering          
         \centering
         \includegraphics[scale=0.52]{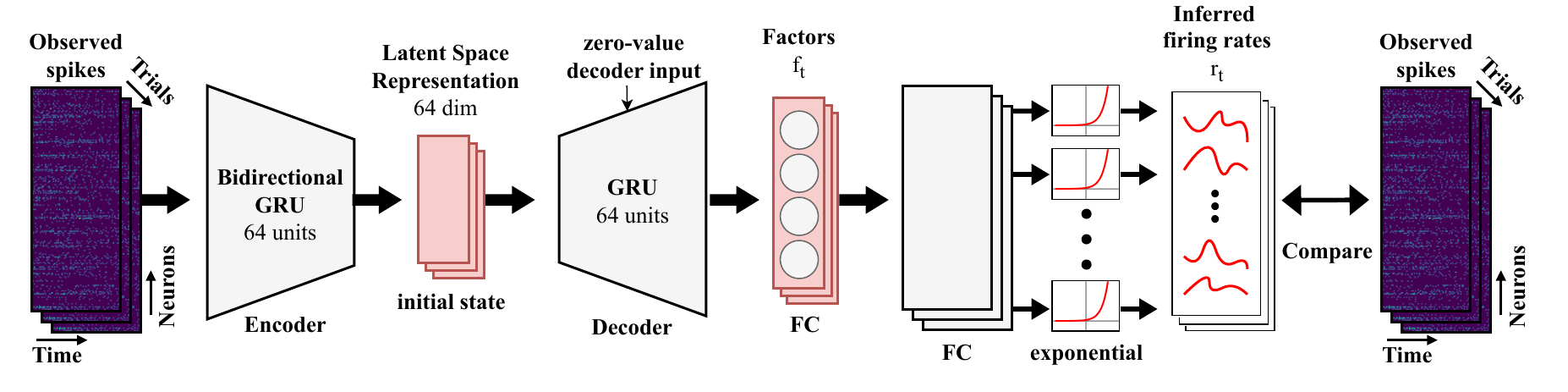}
        \caption{LFADS architecture used for this study. } 
        \label{fig:lfads_architecture}
\end{figure*}

Over the past decade, the ability to record large-scale neural activity has improved dramatically, providing us with new opportunities to study neural population dynamics.
A powerful strategy for analyzing such high-dimensional measurements is to learn the low-dimensional latent dynamics that explain much of the variance in the measurements.
Analysis of neural signals also widely relies on neural networks \cite{AI_Neuroscience}.
LFADS \cite{sussillo2016lfads, pandarinath2018inferring} model 
is a cutting-edge neural network architecture in the field of neural modeling and computational neuroscience. 
LFADS leverages the capacity of recurrent neural networks (RNNs) for uncovering hidden patterns and combines them with variational inference techniques to infer the latent factors that drive the observed data.
Although LFADS demonstrates satisfactory performance in modeling brain signals, processing the extensive neural recordings in real time still poses significant computational challenges. 
Large-scale neural recordings are commonly involved in novel neuroscience experiments \cite{cunningham2014dimensionality}. 
To process the large-scale neural recordings in real-time, Field Programmable Gate Arrays (FPGAs) are used to accelerate the inference. \\

FPGAs enable customized data processing logic and have gained widespread adoption for achieving highly parallel dataflow processing with minimal latency. 
Thus, low-latency, low-power, and high-throughput model inference can be achieved on FPGAs, which makes large-scale real-time neural experiments possible. 
For instance, Low-latency LFADS can be employed to create real-time closed-loop experiments that decode neural activity to manipulate external devices. 
Furthermore, high-throughput machine learning (ML) in neural-related experiments improves the ability to examine extensive neural recordings. 
This capability is increasingly crucial in modern neuroscience experiments, as the integration of large-scale neural recordings has become a commonplace and essential component.

In this paper, we present an efficient implementation of the LFADS model in High-Level Synthesis (HLS) for the \texttt{hls4ml} package \cite{hls4ml_software}.
This HLS implementation will opens the door for wider low-latency and high-throughput applications of the LFADS models for neural sequential data.
Several optimizations for input/output (IO) ports and data access mechanisms are done while implementing the HLS representation of the LFADS model.

\section{Core Concepts}
\label{core_concepts}

As LFADS is one of the promising models for modeling complex brain activities by inferring smooth dynamics based on the collected neural spiking data, it is considered for this study. 
Typically, data collection by the NeuroPort System (Blackrock Microsystems) and neural control and cueing tasks on the Simulink/xPC real-time platform (MathWorks) can be performed within 10 ms \cite{pandarinath2018inferring}. 
The data collection and experimental control can be done much faster by using hardware like FPGAs.
The focus of the work is to demonstrate the ability to run low-latency inference of LFADS-like models on an FPGA.
A modified LFADS model is used for this study as described in Sec. \ref{sec:model}.

\subsection{Model Description}
\label{sec:model}

LFADS is a sequential model based on a variational autoencoder (VAE). 
The encoder has a bidirectional GRU layer, which takes the single-trial neural spikes as input and converts them into a low-dimensional latent space representation. 
This representation acts as the initial state for the decoder GRU layer, which tries to regenerate the input data. 
LFADS assumes the observed spikes are samples from a Poisson process with firing rates ($r_{t}$).
So, instead of estimating the neural spikes, the decoder learns the firing rates ($r_{t}$) as a function of time. 
The decoder (or generator) GRU is expected to learn the underlying dynamics. 
So, the decoder is trained to infer a reduced set of latent dynamic factors, $f_{t}$.
The firing rates can be constructed from the latent factors.\\

In this work, we have studied an autoencoder (AE)-based model to reduce the complexity of FPGA deployment. 
Furthermore, our studies show that removing the Gaussian-sampling has almost no impact on the model performance. 
The original LFADS model uses a controller network to predict the dynamics of the system in the presence of an external input.
To simplify the model, we have studied an LFADS, which can only predict the autonomous dynamics. 
Figure \ref{fig:lfads_architecture} presents the overview of the LFADS architecture used in this work. 
The model structure is adopted from \cite{hurwitz2021targeted}. 
In our model, a bidirectional GRU encoder with 64 units for both forward and backward layers, compresses the input spikes into a latent vector of dimension 64. 
Then the latent variables are used for initializing the unidirectional GRU decoder, which has 64 GRU units. 
The decoder produces a set of low-dimensional temporal factors of dimension 4. 
These four latent dynamic factors are then passed through a fully connected layer with 70 units to produce the log firing rates, $\log(r_t)$, corresponding to the input spike.
The model is trained with Keras \cite{keras} and TensorFlow \cite{tensorflow}.

\subsection{Dataset and Evaluation Metrics}

In this paper, we have used the data collected from a study of monkey reaching tasks \cite{perich2018neural}. 
The input dataset contains a total of 170 trials. 
Out of which, 136 trials (80\%) are used for training, while 17 trials are used for each, validation and test. 
Each trial consists of 70 recording channels, with 73 discrete time steps per channel, resulting a shape of (136,73,70) for the training dataset. 
A detailed training and evaluation methodologies are presented in Ref. \cite{hurwitz2021targeted}. \\

\begin{figure}[!h]
     \centering          
    \includegraphics[width=\linewidth]{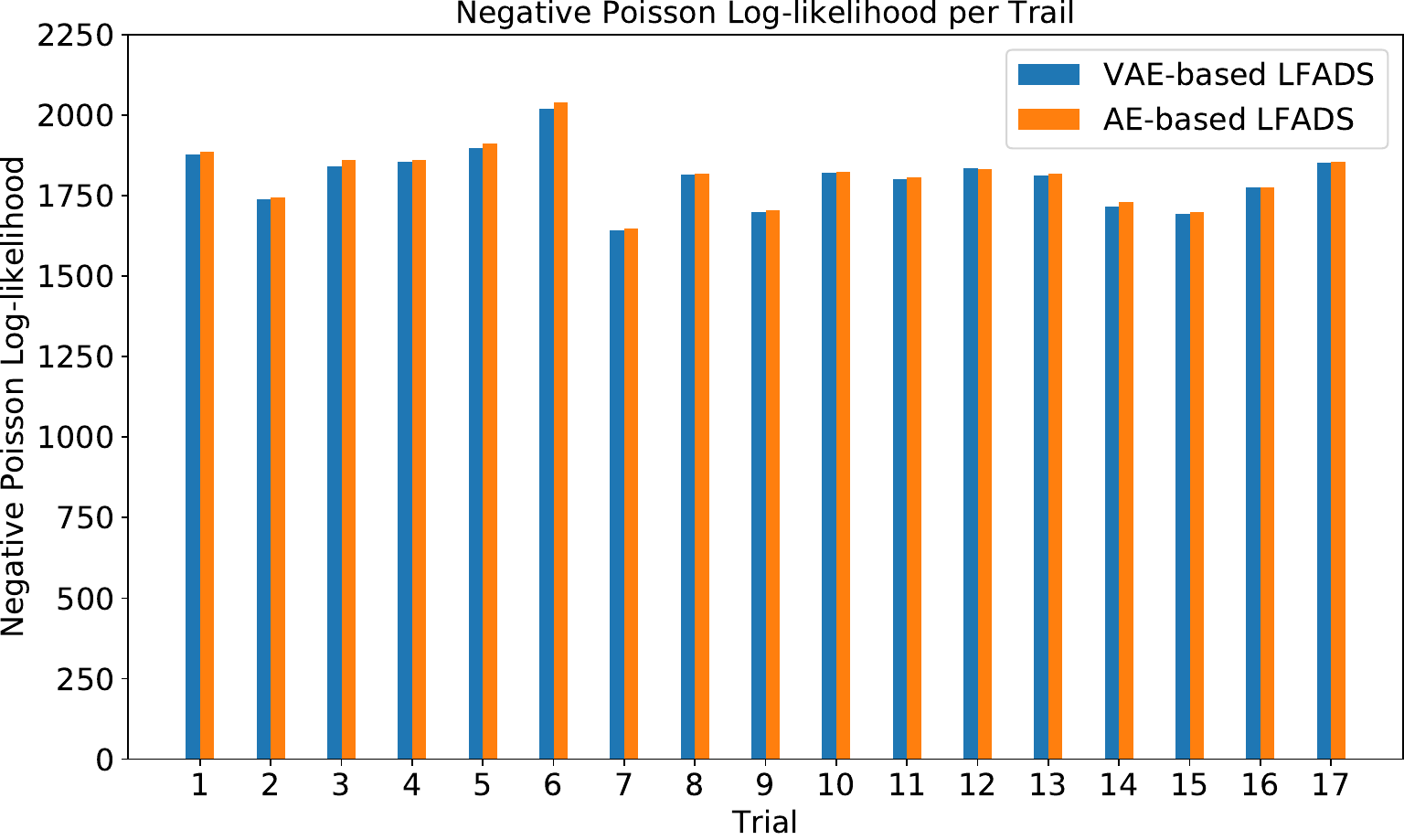}
    \caption{The figure shows the NPLL values of the AE-based LFADS (red) and a VAE-based LFADS (blue) with a similar architecture.}
    \label{fig:non-guassian_compare}
\end{figure}

Two metrics are used to evaluate the performance of LFADS model: negative Poisson log-likelihood (NPLL) and the coefficient of determination (R$^2$).
As LFADS estimates the firing rates, assuming spiking variability follows a Poisson distribution,  it uses NPLL as a loss function in training. 
It should be noted that the main goal of this work is to deploy LFADS onto an FPGA, not to optimize its performance. 
Figure \ref{fig:non-guassian_compare} shows the NPLL comparison of a VAE-based LFADS model with the AE-based model used in this work.
The R$^2$ score is calculated by fitting the reconstructed temporal factors $f_t$ to the measured behavioral data (hand position). 
The closer the score is to 1, the better the factors align with the behavioral data, indicating a stronger correlation between the model's predictions and the observed behavior.

\section{Implementation}
\label{implementation}

One of the main focuses of this work is to implement the LFADS model in HLS and integrate it into \texttt{hls4ml} for rapid development on FPGA. Currently, almost all components of the LFADS model are available in \texttt{hls4ml}, except for the Bidirectional wrapper for the GRU and the quantized GRU (QGRU).

\subsection{HLS implementation: Keras Model}

The HLS bidirectional wrapper is used for the Bidirectional layer of the Keras LFADS model. 
To implement the Bidirectional wrapper, a reverse function is applied to invert the order of the input sequence. 
During inference, the original and reversed inputs pass through corresponding GRU layers and their outputs are concatenated.

\subsection{QKeras Model}

We have also studied quantization-aware training (QAT) using the QKeras package \cite{qkeras} . 
We have built our QKeras model using the QGRU, QBidirectional, QDense layers and different quantized form of sigmoid and tanh activation functions. 
Since these two activation functions are non-linear, outputs of the original and the quantized function could be different. 
So, this has been properly optimized in our implementation. 
For QBidirectional, we use two QGRUs, one for the forward and one for the backward input sequences, and concatenate the two outputs to form the QBidirectional output. 
It is important to note the precision of the inputs also needs to be adjusted by using a quantizer to make different operations consistent.  \\

Most of the hyperparameters used for QAT are similar to the floating-point models. 
For the QKeras model, a step learning rate schedule is used, with an initial learning rate of 0.01, patience of 10 epochs, decay factor of 0.5, and a minimum learning rate of $10^{-5}$. 
The SGD optimizer is used instead of Adam when the total bits are less than 6 bits for more stable training. 
After through scanning, we have chosen to use 3 integer bits for all activations and 1 integer bit (only sign bit) for weights and biases, since they are initialized with the Lecun uniform initializer \cite{lecumuniform}.

\begin{figure}[!h]
     \centering          
         \centering
         \includegraphics[width=\linewidth]{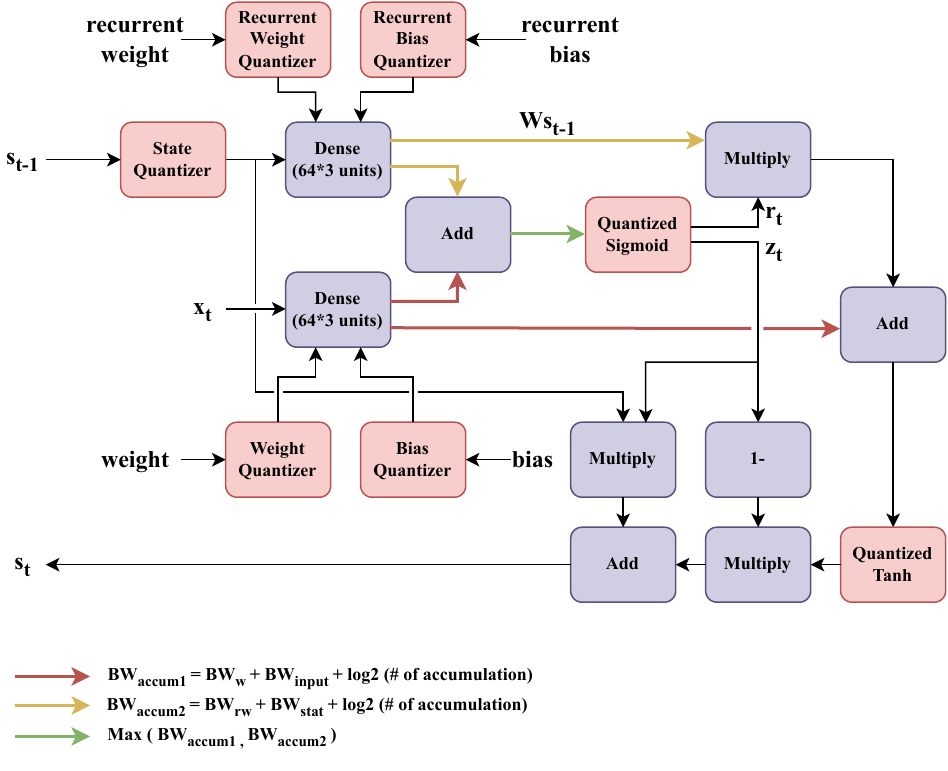}
        \caption{The structure of 64-units quantized GRU cell}
        \label{fig:QGRU_cell}
\end{figure}

\subsection{HLS implementation: QKeras Model}

Figure \ref{fig:QGRU_cell} displays the architecture of a 64-unit quantized GRU cell. There are four weight quantizers, two activation quantizers (quantized sigmoid, and quantized tanh) and one state quantizer.
During the \texttt{hls4ml} compilation, the precisions for all the following calculations are automatically determined based on those quantizers. For example, the colored arrows in Figure \ref{fig:QGRU_cell} show the required bit-width for the addition before the sigmoid. 
\\

\begin{figure}[!h]
     \centering          
     \begin{subfigure}[b]{0.48\linewidth}
         \centering
         \includegraphics[width=\linewidth]{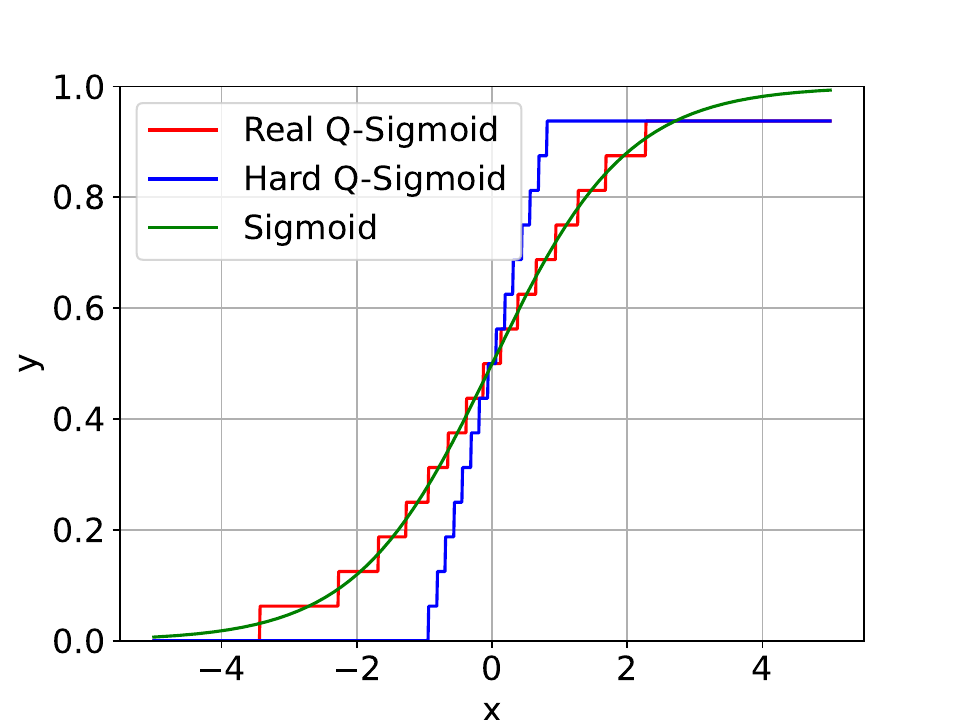}
        \caption{}
        \label{fig:Qsigmoid}
     \end{subfigure}
     \hfill
     \begin{subfigure}[b]{0.48\linewidth}
         \centering
         \includegraphics[width=\linewidth]{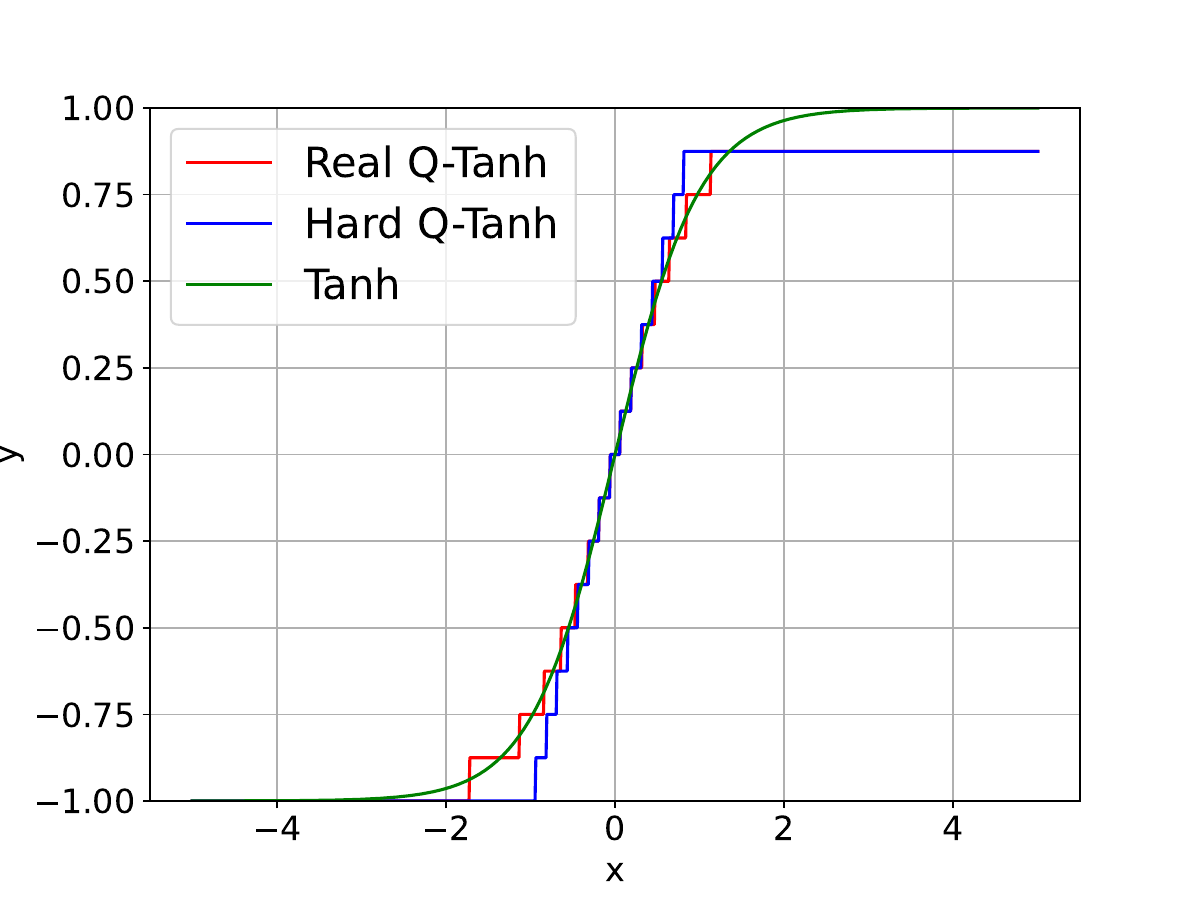}
        \caption{}
        \label{fig:Qtanh}
     \end{subfigure}
\caption{The plots of 4-bits quantized activations, quantized hard activations, and real activations (a) quantized sigmoid and quantized hard sigmoid ranged from 0 to 0.9375. (b) quantized tanh and quantized hard tanh ranged from -1 to 0.875.}
\label{fig:quantized_activation}
\end{figure}

For the choice of quantized activation, we use hard quantized activation. Figure \ref{fig:quantized_activation} shows the curves of the real activations and the quantized (hard) activations. In HLS, it requires a look-up table to achieve the non-linearity in quantized activations. However, as the bit-width increases, the size of the look-up table exponentially grows to maintain accuracy. In contrast, the hard sigmoid calculation can be supported with simple wiring in hardware, without requiring a multiplier or look-up table. This statement is also true for hard tanh as it can be derived from hard sigmoid.

The current version of \texttt{hls4ml} has two options for the datatype used in the dataflow: IO-parallel and IO-stream. 
To maximize the bandwidth, fully-partitioned arrays were utilized in IO-parallel, which can be resource-intensive. 
As a result, we use streams since they are synthesized as more resource-efficient FIFOs. 
Furthermore, we have modified the \texttt{hls4ml} compiler to use array-of streams datatype, offering more flexibility in terms of bandwidth compared to the default \texttt{packed struct} datatype.
The GRU layers only support IO-parallel, so we have applied a simple wrapper layer to the GRU to convert the stream type to an array type for IO. 
Similarly, for the bidirectional layer, we have employed the same approach to handle IO.

\section{Results}
\label{result}

The LFADS model described in Sec. \ref{sec:model} is translated into an HLS model by utilizing the \texttt{hls4ml} tool.
The implementation is tested using Vivado HLS 2020.1 with a Xilinx Alveo U55C FPGA as the target. 
In this work, we have carefully optimized quantization and resource utilization on the FPGA.

\subsection{Quantization Results}
\label{quantization_result}

\begin{figure}[!h]
     \centering          
     \begin{subfigure}[b]{0.48\linewidth}
         \centering
         \includegraphics[width=\linewidth]{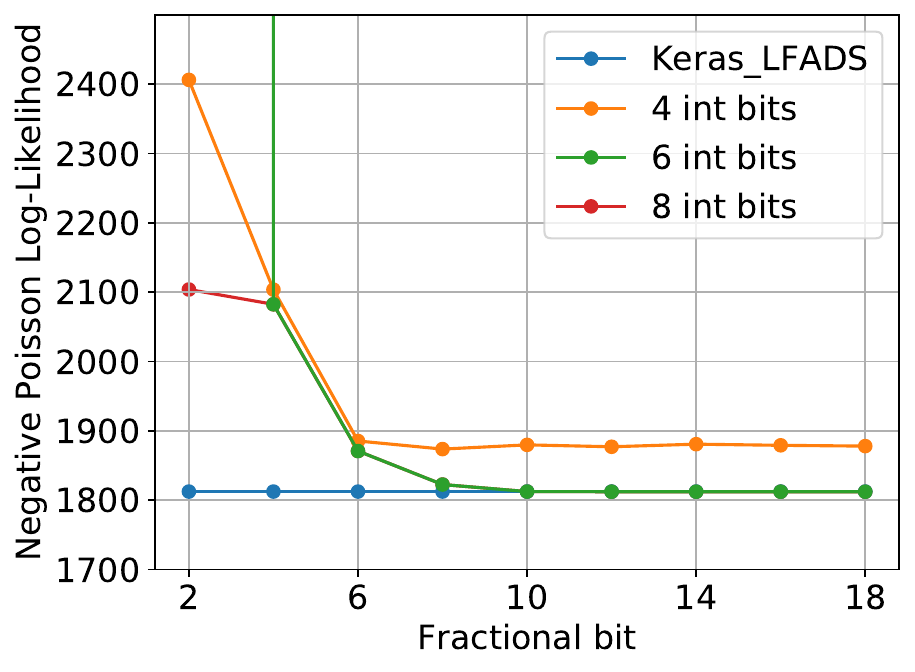}
         \caption{PTQ NPLL}
         \label{fig:PTQ_NPLL}
     \end{subfigure}
     \hfill
     \begin{subfigure}[b]{0.48\linewidth}
         \centering         
         \includegraphics[width=\linewidth]{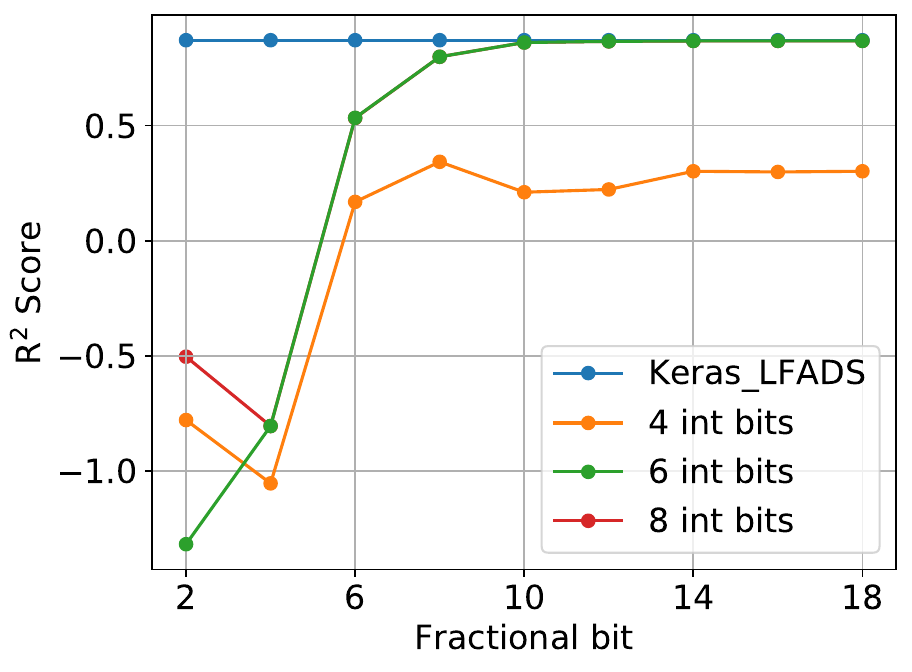}
         \caption{PTQ R$^2$}
         \label{fig:PTQ_R2}
     \end{subfigure}
\caption{Shows (a) NPLL and (b) R$^2$ score as a function of fractional bits. The blue line in each figure represents the floating-point, whereas the lines correspond to inter bits of 4 (orange), 6 (green), or 8 (red).
}
\label{ptq_comparison}
\end{figure}

The quantization process is to reduce the precision of the model parameters as well as the inputs.
Typically the model parameters, such as weights and biases, are stored as 32-bit floating-point numbers in the Keras models. 
As the floating-point number takes up a lot of resources on the FPGA, it is preferred to use fixed-point representations of the model parameters.
The fixed-point format is referred to Vivado HLS \texttt{ap\_fixed} type \cite{hlsapfix}, where the sign bit is included in the integer bits.
In this is paper, the  \texttt{ap\_fixed} numbers are written in \texttt{<total bits, integer bits>} format.
\texttt{hls4ml} converts the input model parameters into \texttt{ap\_fixed} while implementing the HLS model representation. 
This process is called post-training quantization (PTQ).
In our study,  we varied the fractional bits between 2 and 16 while keeping the precision of the integer part fixed to 4, 6, and 8.
The NPLL and $R^2$ scores are calculated for different quantization settings and the values are compared with that of the floating-point model as shown in Figure \ref{ptq_comparison}.
Different colors in the figure represent integer bits of 4 (orange), 6 (green) or 8 (red).
We observe that at least 6 integer bits and 10 fractional bits, \texttt{<16,6>}, are needed to achieve a similar performance as the floating-point model. 

\begin{figure}[!h]
     \centering          
     \begin{subfigure}[b]{0.48\linewidth}
         \centering
         \includegraphics[width=\linewidth]{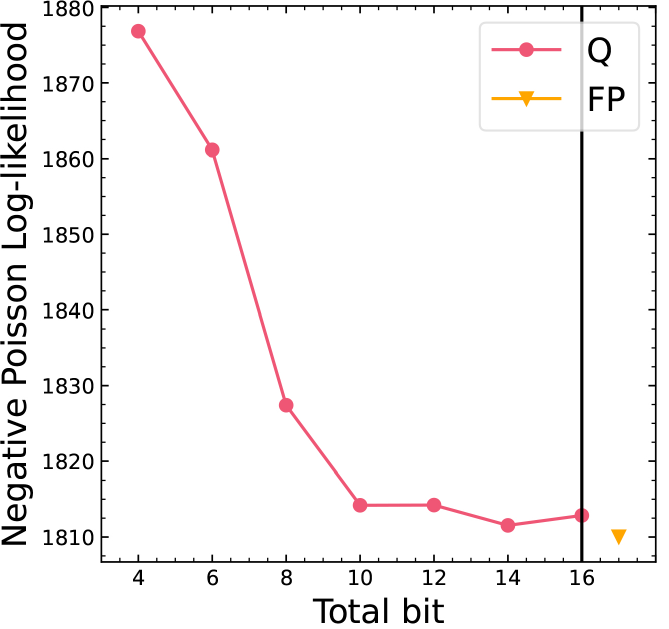}
         \caption{QAT NPLL}
         \label{fig:QAT_NPLL}
     \end{subfigure}
     \hfill
     \begin{subfigure}[b]{0.485\linewidth}
         \centering         
         \includegraphics[width=\linewidth]{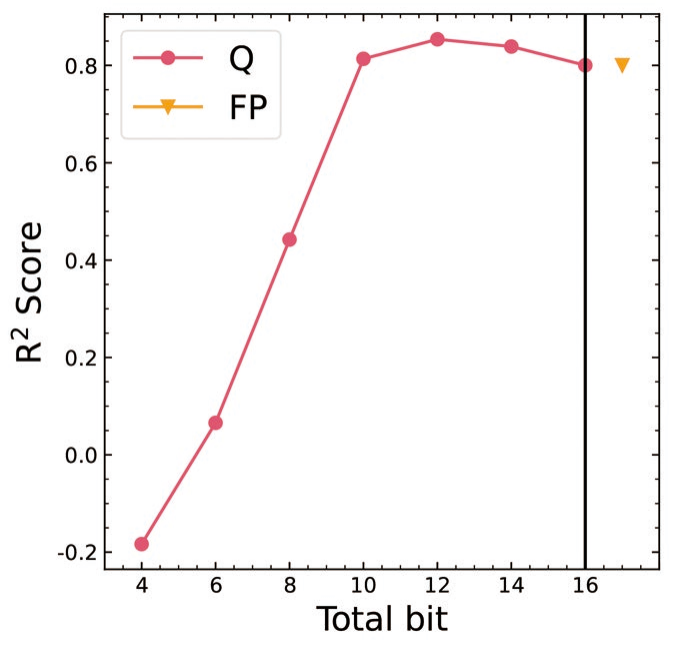}
         \caption{QAT R$^2$}
         \label{fig:QAT_R2}
     \end{subfigure}
\caption{ Shows (a) NPLL and (b) R$^2$ score as a function of total bits for the quantized-aware trained (Q) models using. Results from the baseline floating-point (FP) model are shown in the right subplots.
}
\label{qat_comparison}
\end{figure}

\begin{figure}[!h]
     \centering          
     \begin{subfigure}[b]{0.48\linewidth}
         \centering
         \includegraphics[width=\linewidth]{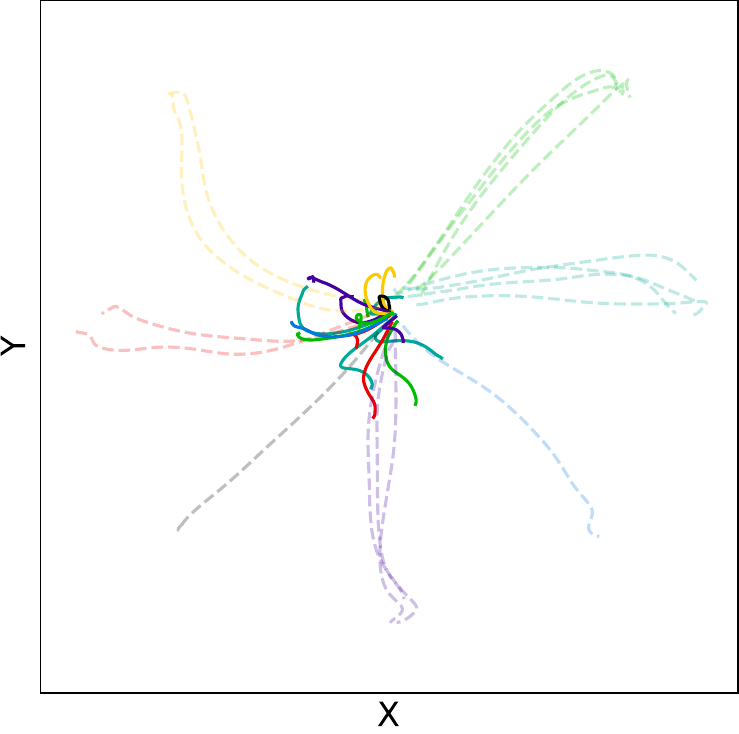}
        \caption{4 bits}
        \label{fig:r2_4}
     \end{subfigure}
    \hfill
     \begin{subfigure}[b]{0.48\linewidth}
         \centering
         \includegraphics[width=\linewidth]{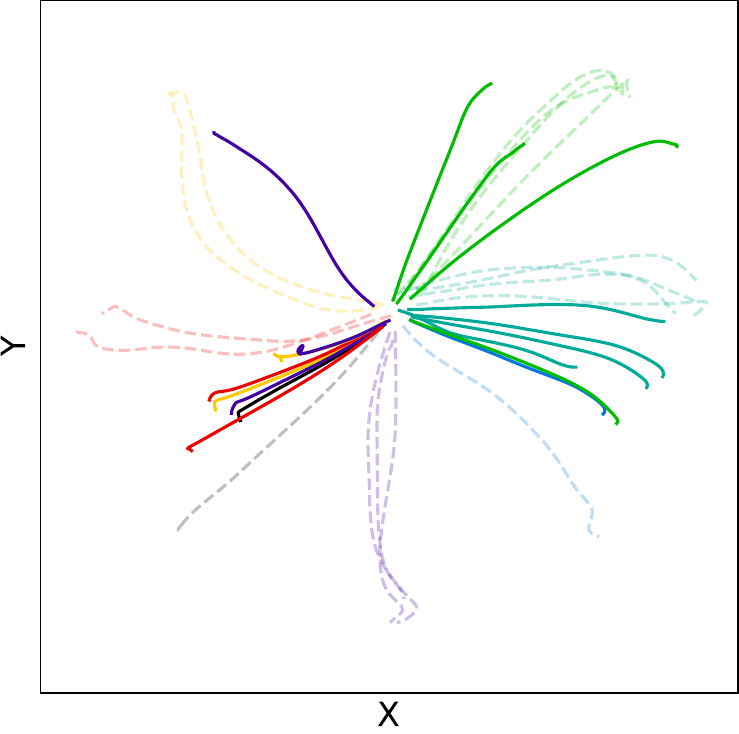}
        \caption{8 bits}
        \label{fig:r2_8}
     \end{subfigure}
    \vfill
     \begin{subfigure}[b]{0.48\linewidth}
         \centering
         \includegraphics[width=\linewidth]{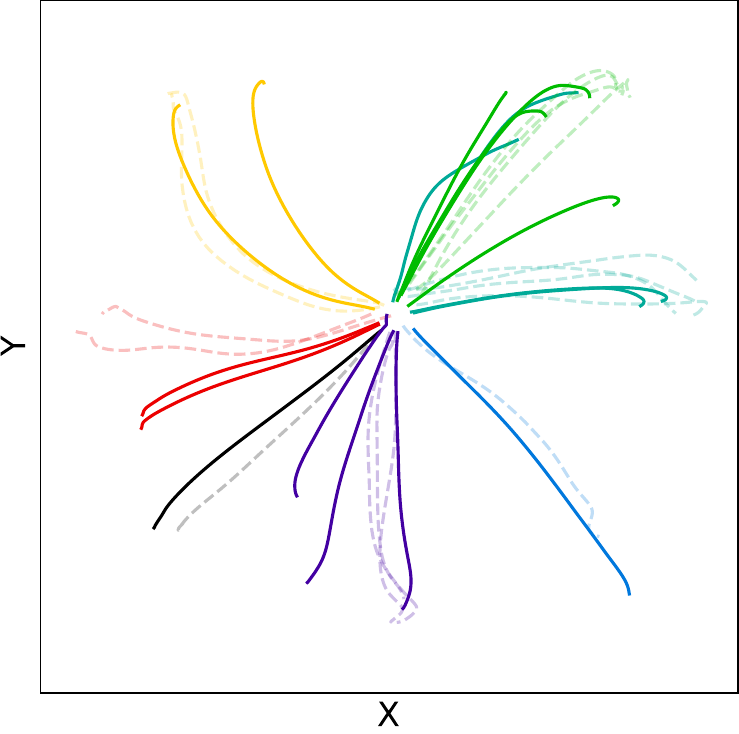}
        \caption{12 bits}
        \label{fig:r2_12}
     \end{subfigure}
         \hfill
     \begin{subfigure}[b]{0.48\linewidth}
         \centering
         \includegraphics[width=\linewidth]{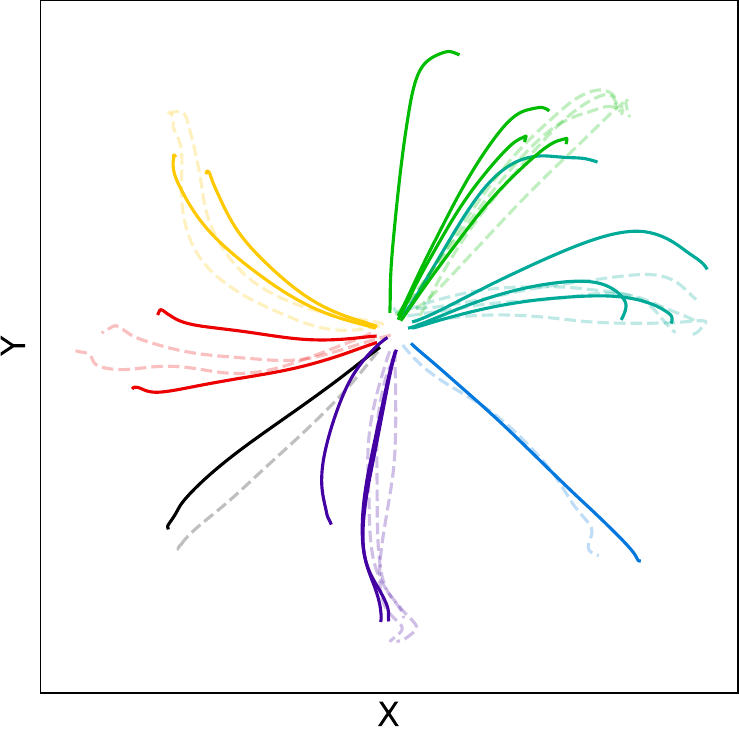}
        \caption{floating point}
        \label{fig:r2_fp}
     \end{subfigure}
\caption{The visualizations compare the reconstructed and target behavior. The reconstructed hand movement trajectories are predicted by the QAT models trained with (a) 4, (b) 8, and (c) 12 total bits and (d) floating-point model with the target behavior. The movements in different directions are grouped and shown in different colors. The solid lines denote the reconstructed movements, while the dotted lines represent the target movements.}
\label{fig:Behaviour_reconstruction}
\end{figure}

The performance of the QAT model is estimated for different total bit widths between 4 and 16, as shown in Figure \ref{qat_comparison}.
We see a noticeable degradation in performance below total width of 10 bits.
A similar observation can be seen in the behavior reconstruction 
shown in Figure \ref{fig:Behaviour_reconstruction}. 
The figure shows the hand movement trajectories in the 2D $x-y$ plan.
The movements performed by the subject in the same direction are grouped together and denoted by the same color. 
The dotted lines represent the target movement direction, where the reconstructed movement trajectories are shown in solid lines.
The QAT models trained with less than 10 total widths show significant degradation in performance.
The performance of the 12-bit QAT model is much closer to the floating-point model.
Based on the results, we have selected the QAT model with 10 bits for subsequent steps.

\subsection{Resource Utilization}

\begin{figure}[!h]
    \centering
    \includegraphics[width=\linewidth]{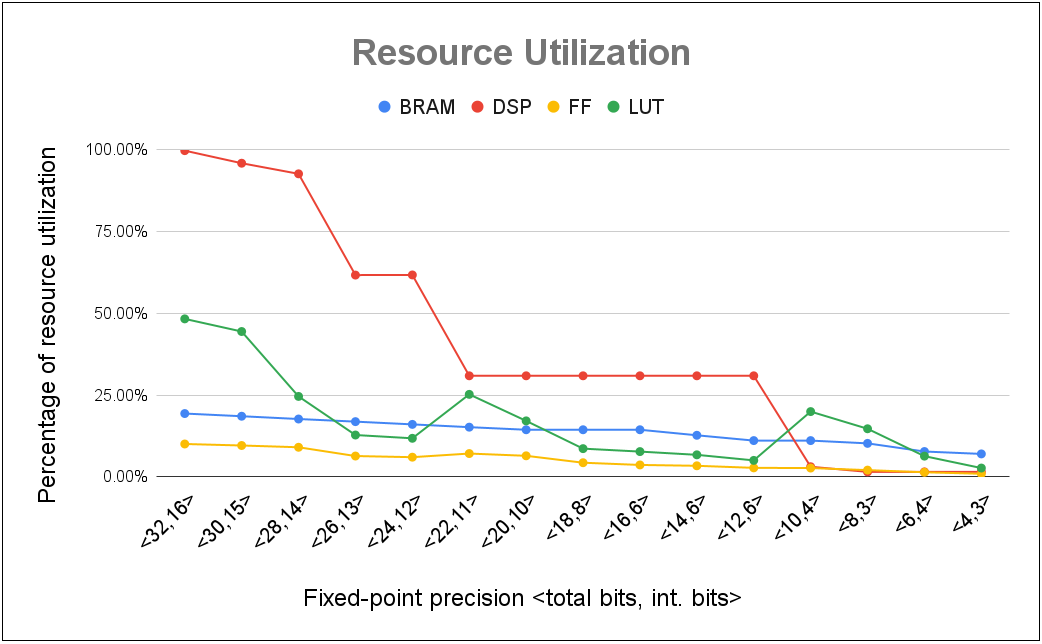}
    \caption{Resource utilization for different total bits with post-training quantization. The target FPGA is Xilinx Alveo U250.}
    \label{fig:Resource_Utilization}
\end{figure}

We have used Alveo U250, which is currently the largest board in the Xilinx Alveo FPGA series, as the target device to ensure the the available resources are sufficient at a high bit width. 
The LFADS model is synthesized using Vivado HLS with different quantization precision to obtain a resource estimation.
For each case, we estimate the utilization of different FPGA resources like memory (BRAM), digital signal processing units (DSPs), flip-flops (FFs), and lookup tables (LUTs).
Figure \ref{fig:Resource_Utilization} shows the resource consumption of the LFADs model with post-training quantization in relation to the total available resources.
The four curves, representing different resources, exhibit a downward trend as the total bit width decreases. 
The limitation of FPGA inference in high bitwidth is DSPs. 
The DSP consumption decreases significantly from 32 bits to 22 bits, followed by a stable trend until 11 bits. 
This is due to the maximum input size for multiplication in DSP48E2 is 27 × 18 \cite{dsp48e2}. 
If an input exceeds this limit, two DSPs will be applied to perform the multiplication. 
For bit widths below 12 bits, the DSP utilization decreases to nearly zero percent since the multiplication is carried out by LUTs, resulting in an increase in LUT consumption from the utilization of 12 bits to 10 bits.

\subsection{FPGA Latency}

To quantify latency, we selected the optimal PTQ model with precision ap\_fixed<16,6>, comprising six integer bits and ten fractional bits.
Given the constraints of our experimental setup, we opted for a more compact FPGA board, the Xilinx Alveo U55C, as it suits our requirements.
Despite its smaller size compared to the Alveo U250, the U55C board comfortably accommodates the best PTQ model, with minimal anticipated alterations in latency results.
Table \ref{tab:deployment} shows the configuration used for this test.
The actual running frequency on the U55C is 200 MHz, and the estimated latency is 41.97 $\mu$s.

\begin{table}[h!]
  \caption{Deployment Configuration}
  \label{tab:deployment}
  \begin{tabular}{cc}
    \toprule
    Target FPGA & Xilinx Alveo U55C\\
    Precision & ap\_fixed<16,6>\\
    Frequency &200 MHz\\
    Latency & 41.97$\mu$s \\
  \bottomrule
\end{tabular}
\end{table}

\section{Summary and Outlook}
\label{conclusion}

We develop an automated design workflow based on \texttt{hls4ml} to deploy LFADS onto an FPGA to accelerate the model for potential large-scale real-time neuroscience experiments. 
For efficient implementation of the model, we studied both PTQ and QAT.
Using QAT it is possible to reduce total number of bits required to 10 bits, compared to the 16 bits required for PTQ, while incurring negligible loss compared to the floating-point model.
For the on-board evaluation, the 16-bit model demonstrates a latency of 41.97 $\mu$s for processing the data in a single trial, which is relatively small compared to the $\mathcal{O}(ms)$ sampling interval commonly used in neuroscience experiments. The acceleration enables large-scale real-time experiments and enhances the capacity to process extensive neural recordings. 
An AE-based LFADS model is studied for this work. 
The VAE-based LFADS should be supported by \texttt{hls4ml} in the near future.

\begin{acks}
We acknowledge the Fast Machine Learning collective and NSF A3D3 team as an open community of multi-domain experts and collaborators.
This communities were important for the development of this project.
We thank Eli Shlizerman, Amy Orsborn, Cole Hurwitz, and Michael Nolan for valuable discussions at different stages of the project. 
This research was funded in part by National Science Foundation (NSF) grants No. 1934360 and 2117997, and 1934360.

\end{acks}

\section*{CODE AVAILABILITY STATEMENT}
We re-implemented the LFADS in Keras functional API. The model can be found in the "Tensorflow 2 Keras LFADS" repository available at \url{https://github.com/XLSMALL/TF2Keras-LFADs}, originally developed by Hurwitz et al. The original work can be found at \url{https://github.com/HennigLab/tndm}. The PTQ model conversion was done using the \texttt{hls4ml} package found at \url{https://github.com/XLSMALL/hls4ml/tree/LFADs}, and we intend to contribute the QAT code to \texttt{hls4ml} in the near future. 


\bibliographystyle{ACM-Reference-Format}
\bibliography{ref.bib}

\end{document}